\documentclass[aps,physrev,preprint,superscriptaddress,nofootinbib]{revtex4-2}

\usepackage{psfrag,slashed,cancel,array,graphicx}
\usepackage[utf8]{inputenc}
\usepackage{mathtools}
\usepackage{mathrsfs}
\usepackage{multirow}
\usepackage{braket}
\usepackage{slashed}
\usepackage{booktabs}
\usepackage{slashed}
\usepackage{hyperref}
\usepackage{cleveref}
\usepackage{amssymb}

\hypersetup{pdftitle={},pdfcreator={},linkcolor=[rgb]{0.15,0.35,0.75},colorlinks=true,citecolor=[rgb]{0.675,0,0.2},urlcolor=[rgb]{0.15,0.35,0.65}}

\def\beq{\begin{equation}}
\def\eeq{\end{equation}}
\def\bsp#1\esp{\begin{split}#1\end{split}}
\def\d{{\rm d}}
\long\def\bea#1\eea{\begin{align}#1\end{align}}

\newcommand{\nn}{\nonumber}

\AtBeginDocument{
\heavyrulewidth=.08em
\lightrulewidth=.05em
\cmidrulewidth=.03em
\belowrulesep=.65ex
\belowbottomsep=0pt
\aboverulesep=.4ex
\abovetopsep=0pt
\cmidrulesep=\doublerulesep
\cmidrulekern=.5em
\defaultaddspace=.5em
}

\def\be{\begin{equation}}
\def\ee{\end{equation}}

\begin{document}

\title{Tracing Vacuum Hadronization with Conserved Currents }

\author{Weiyao Ke}
\email{weiyaoke@ccnu.edu.cn}
\affiliation{Key Laboratory of Quark and Lepton Physics (MOE) \& Institute of Particle Physics, Central China Normal University, Wuhan, Hubei 430079, China}
\affiliation{Southern Center for Nuclear-Science Theory (SCNT), Institute of Modern Physics, Chinese Academy of Sciences, Huizhou, Guangdong 516000, China}

\author{Hai Tao Li}
\email{haitao.li@sdu.edu.cn}
\affiliation{School of Physics, Shandong University, Jinan, Shandong 250100, China}

\author{Wanchen Li}
\email{wanchenli@fudan.edu.cn}
\affiliation{Department of Physics, Center for Field Theory and Particle Physics, and Key Laboratory of Nuclear Physics and Ion-beam Application (MOE), Fudan University, Shanghai, 200433, China}

\author{Xiaohui Liu}
\email{xiliu@bnu.edu.cn}
\affiliation{School of Physics and Astronomy, Beijing Normal University, and Key Laboratory of Multiscale Spin Physics (Beijing Normal University), Ministry of Education, Beijing 100875, China}
\affiliation{Southern Center for Nuclear-Science Theory (SCNT), Institute of Modern Physics, Chinese Academy of Sciences, Huizhou, Guangdong 516000, China}

\author{Ding Yu Shao}
\email{dyshao@fudan.edu.cn}
\affiliation{Department of Physics, Center for Field Theory and Particle Physics, and Key Laboratory of Nuclear Physics and Ion-beam Application (MOE), Fudan University, Shanghai, 200433, China}
\affiliation{Shanghai Research Center for Theoretical Nuclear Physics, NSFC and Fudan University, Shanghai 200438, China}
\affiliation{Center for High Energy Physics, Peking University, Beijing 100871, China}
\affiliation{Southern Center for Nuclear-Science Theory (SCNT), Institute of Modern Physics, Chinese Academy of Sciences, Huizhou, Guangdong 516000, China}

\begin{abstract}
We show that color triality constrains the nonperturbative states that screen a Wilson-line endpoint, allowing the joint flows of net electric charge, baryon number, and strangeness to probe QCD vacuum hadronization. Whether evaluated from resolved hadrons in a jet initiated by a quark of flavor $f$ or from the corresponding charge correlators, these flows satisfy the Gell-Mann--Nishijima relation 
\bea
\langle Q_{\rm flow}\rangle_f \simeq I_{3,f} +\frac{1}{2}\left( \langle S_{\rm flow}\rangle_f +\langle B_{\rm flow}\rangle_f \right), \nn 
\eea
where $I_{3,f}$ is the third component of the initiating-quark isospin. The net jet baryon number, $\langle B_{\rm flow}\rangle_f\simeq \frac{r_{qq}}{1+r_{qq}}$, directly probes the relative probability of a diquark--antidiquark vacuum excitation, with $r_{qq}$ the diquark-to-quark production ratio. Thus, for $r_{qq}\ll1$, the baryon number carried by the jet is substantially suppressed relative to the initiating-quark value $B_f=1/3$. Likewise, the strange-to-light pair-production ratio, defined by $u\bar u:d\bar d:s\bar s=1:1:r_s$, is encoded in the measured jet strangeness $\langle S_{\rm flow}\rangle_f \simeq S_f+\frac{3r_s}{2+r_s} \left(\frac{1}{3}- \langle B_{\rm flow}\rangle_f\right)$, predicting a nonzero mean net strangeness even in $u$- and $d$-initiated jets. The conserved-current moments appearing in these relations are independent of the renormalization scale. Their simultaneous measurement therefore provides a direct, flavor-resolved probe of vacuum pair production and quantum-number transport during hadronization.
\end{abstract}

\maketitle
\medskip

\section{Introduction}\label{sec:intro}
Colored partons produced in a QCD hard process~\cite{Gross:1973id,Politzer:1973fx,Collins:1989gx} must ultimately form color-singlet hadrons through hadronization, long-distance QCD dynamics that remains poorly understood~\cite{Webber:1999ui}. Fragmentation functions provide the standard description of inclusive identified-hadron production~\cite{Metz:2016swz}, but do not directly reveal how color is screened, flavor is produced, or baryons are formed. 
Phenomenological string and cluster hadronization models implement these microscopic mechanisms through explicit color-screening degrees of freedom~\cite{Andersson:1983ia,Webber:1983if}, whose production is controlled by parameters such as the relative weights of $s\bar{s}$ to light-quark pairs and of diquark--antidiquark to quark--antiquark pairs $q{\bar q}$~\cite{Bierlich:2022pfr,Gieseke:2017clv,Campbell:2022qmc}. Direct experimental constraints on these hadronization parameters remain challenging, hindering discrimination among hadronization mechanisms.

Here we show that joint measurements of the net electric charge $Q$, baryon number $B$, and strangeness $S$ flows provide a direct probe of microscopic hadronization. For any conserved charge ${\cal Q}=Q\,, B\,, S$, we denote its flow into a region $\mathcal R$ as ${\cal Q}_{\mathcal R}\equiv\sum_{h\in \mathcal R} \mathcal Q_h$, where $\mathcal R$ can be a jet, a hemisphere, or a collinear sector selected by a global event shape.
Unlike the momentum-weighted jet charge $\sum_{h\in {\mathcal R}}z_h^\kappa Q_h$~\cite{Krohn:2012fg,Waalewijn:2012sv}, which is sensitive to momentum sharing, ${\cal Q}_{\cal R}$ only measures the net charge and its moment $\langle {\cal Q}_{\cal R}\rangle $ is infrared safe~\cite{Riembau:2024tom,Cao:2026fzq,Zhang:2026emt}. Once its calculable perturbative effects are separated, any shift from the initiating-parton charge therefore isolates nonperturbative charge transfer during hadronization.

The factorization framework makes this separation precise. A gauge-invariant collinear sector, such as the fragmentation function or a jet, is completed by a Wilson line extending to infinity which potentially prevents the physical hadrons assigned to that sector from forming a closed system with respect to global quantum numbers. Indeed, a quark that initiates the sector carries fractional charges, whereas every asymptotic hadron carries integer quantum numbers. Collins and Rogers~\cite{Collins:2023cuo} proposed that the missing quantum numbers reside in a nonperturbative state bound to the Wilson line at vanishing momentum fraction. 

In this note, we formulate this state as a defect sector ($D$) carrying boundary charges at the Wilson-line puncture. Since the defect charges vanish perturbatively,  the shifts from the initiating-parton charges 
$ \langle {\mathcal Q}_{D} \rangle \equiv \mathcal Q_{\mathrm{parton}} -\big\langle{\mathcal Q}_{\mathcal R}\big\rangle$
directly measure the quantum numbers carried by the nonperturbative state. By recognizing the color-triality~\cite{Greensite:2003bk} selection rule imposed by confinement, we turn this picture into a predictive framework for classifying the allowed nonperturbative states and their global quantum numbers $\langle {\mathcal Q}_{D} \rangle$. Although the Wilson line is itself neutral under the global symmetries, its endpoint carries nonzero color triality and must be screened by dynamical matter of conjugate triality, sharply restricting the quark content and the quantum numbers of the screening state.

On this basis, we establish invertible relations between the joint $Q$, $B$, and $S$ shifts and the underlying vacuum-hadronization probabilities. The net baryon flow determines the vacuum diquark-to-quark excitation ratio $r_{qq}$, while the joint flows determine the strange-to-light ratio $r_s$. These relations predict a nearly vanishing net jet baryon number for $r_{qq}\ll 1$, despite $B_f=1/3$ of the initiating quark, and a nonzero net strangeness in $u$- and $d$-initiated jets. More generally, QCD isospin symmetry of the screening state yields a jet-level Gell-Mann--Nishijima relation among the conserved flows, providing a closure test that requires fewer assumptions than the probability extractions. Pythia and Herwig simulations reproduce the predicted correlations within their respective hadronization models and allow us to extract effective strange- and baryonic screening parameters. This demonstrates that the charge shifts provide information complementary to inclusive hadron-yield fits.
   
The remainder of this note is organized as follows. We first define the charge-flow operator associated with a collinear sector and the complementary defect charge required by charge conservation. We then show how the Wilson line completing the collinear sector produces a puncture on the celestial sphere, thereby providing a geometric interpretation of the defect sector. Applying the color-triality selection rule together with a minimal, generic assumption about color screening, we classify the electric charge, strangeness, and baryon number that can be carried by the screening state and derive the resulting nonperturbative shifts to the observed charges, which are renormalization-scale independent at leading power. These shifts are inverted to extract the vacuum-screening probabilities $r_s$ and $r_{qq}$, leading to the jet Gell-Mann--Nishijima closure relation. Using \textsc{Pythia}~8 and \textsc{Herwig}~7.3, we validate the jet Gell-Mann--Nishijima relation and the extraction of $r_s$ and $r_{qq}$. The appendices extend these tests to correlators.

\section{Brief Review of Factorization }\label{sec:factorization}
For brevity, we refer to both a jet with its net electric charge measured and a charge-flow correlation event shape as a charge-measured jet. Throughout the note, we assume that the jet is sufficiently inclusive to contain the entire collinear radiation. The cross section factorizes schematically as
\bea\label{eq:fact}
d\sigma[{Q}_{\rm flow}] \sim f_a\otimes f_b\otimes \hat{\sigma}_{ab\to f} \, J_f[{Q}_{\rm flow}],
\eea
where $f_{a,b}$ are parton distribution functions for hadronic collisions and $f_{a/b}=1$ when a hadron beam is replaced by an electron beam. The partonic cross section $\hat \sigma_{ab\to f}$ itself admits a refactorization into the hard, soft functions and collinear matching coefficient associated with the $f$-quark jet. The charge-measured jet function is, at leading power~\cite{Li:2026fhd},
\bea\label{eq:jetQ}
J_f[{Q}_{\rm flow}] = \frac{1}{2N_C}\int d\xi d^2 \boldsymbol x_T \, e^{i\omega\xi} {\rm Tr} \left[ \frac{\gamma^+}{2} \langle0|\xi_{n,f}(\xi, \boldsymbol x_T) W_n^\dagger \,\widehat{Q}_{\rm flow}\,W_n \bar \xi_{n,f}|0\rangle \right] \equiv \langle Q_{\rm flow} \rangle_f  \,,
\eea
Here $\xi_{n,f}$ is the $n$-collinear quark field of flavor $f$ that initiates the jet, $W_n$ is the collinear Wilson line extending in the conjugate $\bar n$ direction, and $\widehat{Q}_{\rm flow}$ measures the net electric charge carried by resolved hadrons of the jet. 

\section{Collinear Charge Flow Operator}\label{sec:chargeflow}
At leading power~\footnote{The angular coverage refers to the factorized $n$-collinear sector, rather than to the full physical event. The collinear radiation selected by a physical region $\mathcal R$ is mapped, in the infinite-boost collinear limit, onto the full celestial space of the $n$-collinear theory. Its non-collinear complement is represented by the eikonal Wilson line along $\bar n$, whose endpoint appears as the excluded puncture $D_\epsilon(\bar n)$. Thus the integral over $S^2\setminus D_\epsilon(\bar n)$ is the collinear representation of the physical hemisphere flow, not a measurement over the full event with one angular point removed.},  
\bea 
\widehat{Q}_{\rm flow} = \lim_{\epsilon \to 0 } \, \int_{S^2\setminus D_\epsilon(\bar n)} d\Omega \, \widehat{{Q}}(\Omega)  \,, 
\eea 
where $\widehat{{Q}}(\Omega)  	=	\lim_{r\to\infty} r^2	\int_{-\infty}^{\infty} \d t\,	\hat n_i J^i(t,r\hat{n}),$ is the electric charge-flow  operator~\cite{Hofman:2008ar,Monni:2025zyv,Cao:2026fzq}, with $J^\mu$ the electromagnetic current and $\hat n$ the direction specified by $\Omega$. Here $D_\epsilon(\bar n)$ is a small angular disk surrounding the antipodal direction $\bar n$ with opening angle $\epsilon$. The $n$-collinear representation uses the stereographic coordinate $\boldsymbol n_T\sim\boldsymbol q_T/\bar n\!\cdot q$~\cite{Li:2026fhd}, where $q^\mu$ denotes an $n$-collinear momentum. This coordinate covers $S^2$ except the point $\bar n$, becoming singular as $\bar n\!\cdot q\to0$. The charge-flow detector is therefore defined on $S^2\setminus D_\epsilon(\bar n)$, while the excluded point $\bar n$ is the celestial endpoint of the Wilson line. 

Note that the limit $\epsilon\to0$ is taken only after separating any puncture-supported contribution from the charge measured by the detector; therefore, the charge flow operator $\widehat{Q}_{\rm flow}$ need not measure the total net charge of the collinear sector. A fraction of the charge quantum number may be carried away by degrees of freedom localized at the excluded puncture $D_\epsilon(\bar n)$, and hence
\bea\label{eq:QflowExp}
\langle Q_{\rm flow}\rangle_f = Q_f- \langle Q_D \rangle_f\,,
\eea 
where $Q_f$ is the conserved total electric charge  of the $n$-collinear sector, and $\langle Q_D \rangle_f $ denotes the mean charge localized at the puncture that is missed by the charge flow measurement. 

\section{Baryon Number and Strangeness Measurement}\label{sec:baryonflow}
An identical construction applies to $\langle B_{\rm flow}\rangle_f$, $\langle B_D\rangle_f$, as well as to $\langle S_{\rm flow}\rangle_f$ and $\langle S_D\rangle_f$, with $\widehat{Q}_{\rm flow}$ replaced by the baryon--flow operator $\widehat{B}_{\rm flow}$ and the electromagnetic current by the conserved baryon current, or the corresponding operators for strangeness. For definiteness, we formulate the following discussion in terms of electric charge. All statements carry over to baryon number and strangeness under these replacements. 

\section{Collinear Wilson Line and the Defect Sector}\label{sec:wilsonline}
We now discuss a possible mechanism for generating a nonzero $\langle Q_D\rangle_f$ or $\langle B_D\rangle_f$ by examining the degrees of freedom associated with the Wilson-line endpoint. Since the excluded puncture is the celestial endpoint of the collinear Wilson line, quantum numbers not measured by the detector may be carried by its associated line-defect sector. 

\begin{figure}[!htb]
    \centering
    \includegraphics[width=0.5\linewidth]{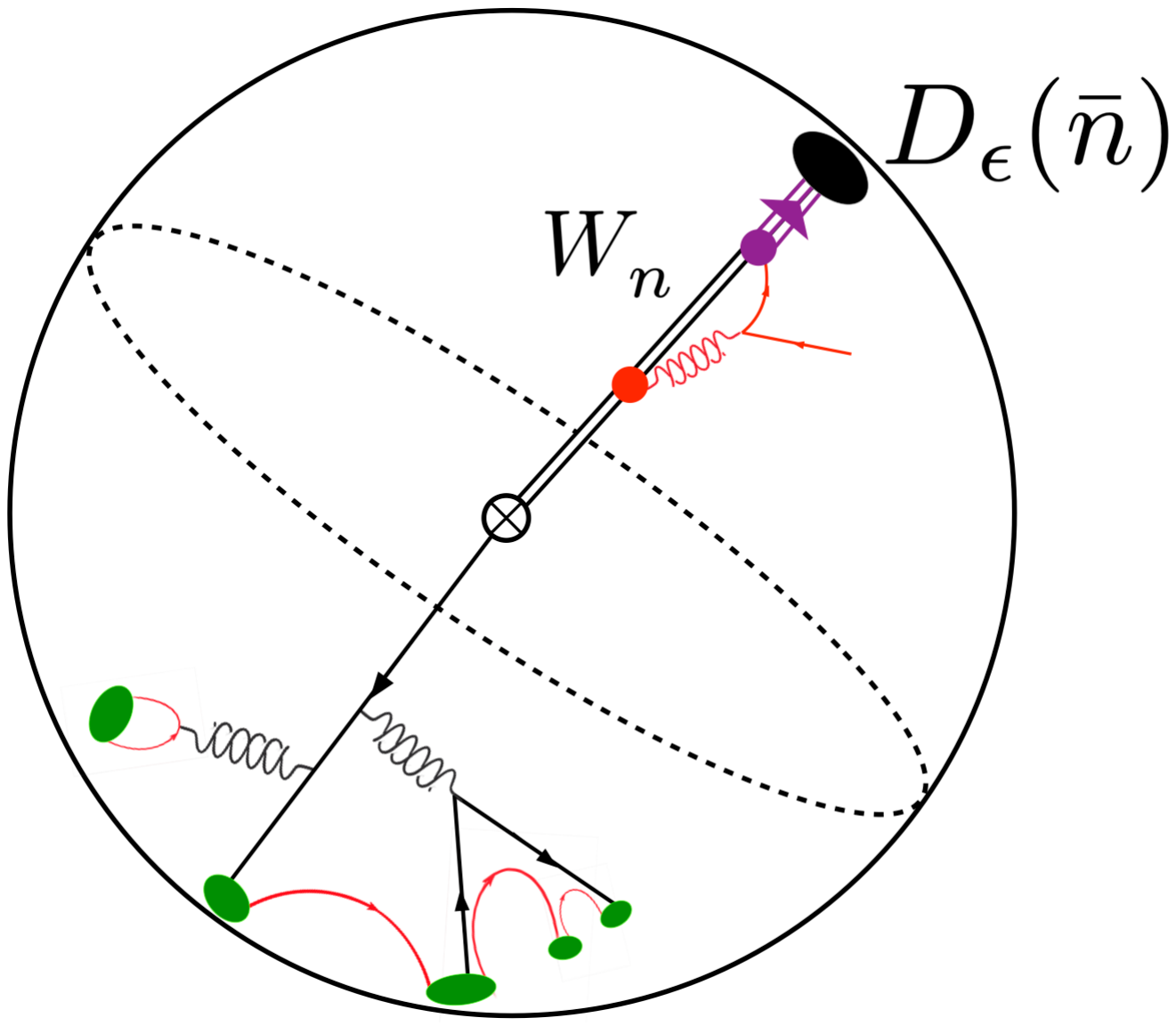}
    \caption{The coverage of $\widehat{Q}_{\rm flow}$ in the collinear sector. The collinear Wilson line defines a line defect and its associated defect sector in the $n$-collinear theory, leaving an area $D_\epsilon(\bar n)$ on the celestial sphere of the collinear theory that is not covered by the charge flow detector. The purple insertion schematically denotes charged defect operators generated through the  OPE of a quark approaching the Wilson line, $ W_n^\dagger \, q(x) \xrightarrow{x_\perp\to0} \sum_i C_i(x_\perp) \mathcal O_{D,i}$. The operators contribute to the puncture charge $Q_D$. Consequently, the charge detector $\widehat {Q}_{\mathrm{flow}}$, which covers $S^2\setminus D_\epsilon(\bar n)$ of the collinear theory, measures only the complementary charge $\langle Q_{\mathrm{flow}} \rangle_f =Q_f-\langle Q_D\rangle_f$. At leading power, $\langle Q_{\mathrm{flow}}\rangle_f$ and $\langle Q_D \rangle_f $ are scale independent. }
    \label{fig:defect}
\end{figure}

The future-pointing collinear Wilson line represents an external eikonal color source and therefore defines a line defect~\cite{Brennan:2022tyl,Brennan:2025acl} of the $n$-collinear theory and its associated defect sector. The line defect extends geometrically along the conjugate direction $\bar n$ and terminates at future null infinity on the celestial sphere, as illustrated in Fig.~\ref{fig:defect}. The defect sector consists of the defect line together with the QCD excitations that can dress it. The sector can be made explicit by representing $W_n$ as the propagator of an auxiliary field $\eta$, with action $S_\eta=\int d\tau\,\bar\eta\,i\bar n\!\cdot D\,\eta$. Together with its QCD dressings, e.g., the charged operators $\bar\eta_i q^i$, $C^{ijk} \bar\eta_i {\bar q}_j {\bar q}_k\,, \dots$, they realize the defect sector. 

Since the puncture location $\theta=\pi$ corresponds to $\bar n\!\cdot q=0$, every ordinary emission with $\bar n\!\cdot q>0$ lies at a finite $\boldsymbol n_T$ and is eventually included in $\widehat {Q}_{\rm flow}$. Consequently, a finite $\langle Q_D\rangle_f$ requires charge-weighted support localized at $\bar n\!\cdot q=0$ that survives the shrinking-disk limit; any locally integrable contribution around the puncture instead vanishes with $D_\epsilon(\bar n)$. In the factorized \(n\)-collinear sector, the Wilson line is the only operator extending to this point, so such endpoint support can arise from excitations that dress or bind to the Wilson line. 

In perturbation theory, pure virtual corrections cannot change the average charge leaking through the puncture, since all QCD interaction insertions are electrically neutral. Real radiation can accumulate at ${\bar n}\cdot q=0$, for instance, via eikonal singular couplings to gluons, and through $g\to q{\bar q}$, it can bring an anti-quark field onto the Wilson line to form a charged defect operator, and thereby leave the charge flow detector, see Fig.~\ref{fig:defect}.  The associated charge-weighted contribution cancels, since for every configuration in which a soft quark enters the puncture, there is an equally weighted configuration in which the corresponding antiquark enters it~\cite{Monni:2025zyv}. More precisely, in perturbation theory, charge weighting projects the radiation onto the charge-odd flavor-nonsinglet channel, whose Regge behavior has no nonintegrable $1/{\bar n}\cdot q$ singularity~\cite{Gehrmann:2026qbl,Mitov:2006ic}. Hence the charge over a small range $\bar n \cdot q < \epsilon$ vanishes as $\epsilon\to0$, so perturbative radiation cannot deposit finite charge around $\bar n \cdot q = 0 $. Consequently,
\bea\label{eq:ward-pert}
\langle Q_D\rangle_f^{{\rm pert}}=0\,,
\qquad \qquad \langle Q_{\rm flow}\rangle_f^{\rm pert}= Q_f \,, 
\eea
at leading power.

\section{Vacuum Hadronization and the Jet Gell-Mann--Nishijima Relation}\label{sec:hadronization}
\begin{figure}[!htb]
    \centering
    \includegraphics[width=0.8\linewidth]{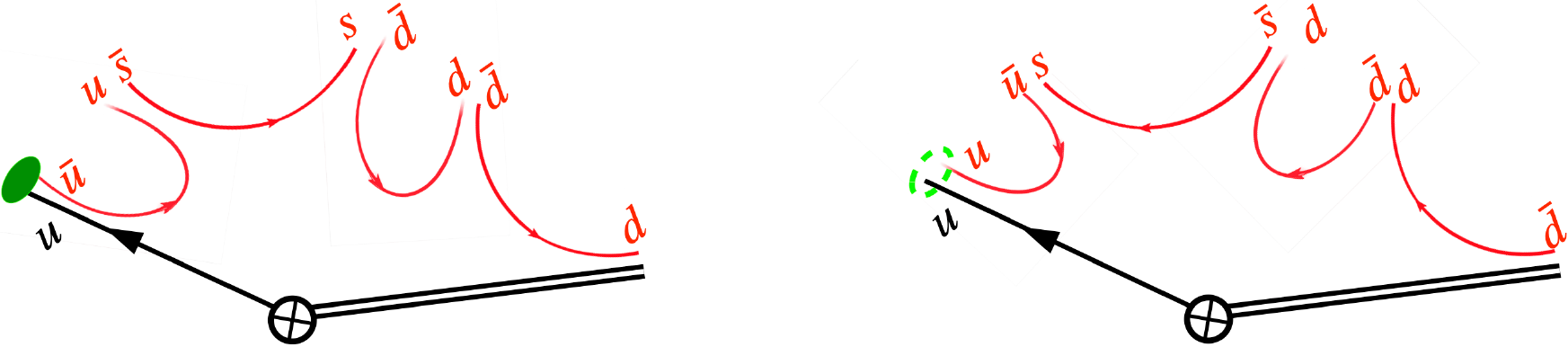}
    \caption{ Hadronization may induce a long-range correlation between the initiating quark, shown in black, and the ordering of soft quark-antiquark pairs along the fragmentation chain, shown in red. In the meson-dominated configurations illustrated in~\cite{Collins:2023cuo}, the initiating $u$ quark can combine with a neighboring $\bar u$, as indicated by the solid green ellipse, whereas the reversed ordering with a neighboring $u$ is suppressed, as indicated by the dashed ellipse. The two orderings leave, respectively, a $d$ and a $\bar d$ onto the Wilson line, contributing $Q_d$ and $-Q_d$ to charges carried by charged defect operators. A probability imbalance between these configurations therefore produces a nonvanishing average $ \langle Q_D \rangle_u$ dominated by $Q_d$, and analogously a nonvanishing boundary baryon number $\langle B_D \rangle_u$.}
    \label{fig:hadronization}
\end{figure}

Nevertheless, non-perturbatively, hadronization might generate strong correlations between the soft quark $q$ distribution and the charged collinear endpoint ${\bar \xi}_{n,f}$, see Fig.~\ref{fig:hadronization} for one example. The conditional distribution in a fixed quark sector $f$ need not retain the $q\leftrightarrow\bar q$ symmetry responsible for the perturbative cancellation. This does not imply that charge conjugation is broken by QCD, since charge conjugation maps the complete quark sector $f$ into the corresponding antiquark sector $\bar f$ and requires only
\bea
\langle Q_D\rangle _{f}=-\langle Q_D \rangle_{\bar f}\,.
\eea
The puncture can therefore acquire a non-vanishing non-perturbative charge expectation value,
\bea
\langle Q_D \rangle_f \neq0\,,
\qquad \qquad \langle Q_{\rm flow}\rangle_f =Q_f- \langle Q_D \rangle_f \,.
\eea
The value of $\langle Q_D \rangle_f$, whether $0$ or not, is determined by the QCD non-perturbative dynamics, such as the string fragmentation mechanism~\cite{Webber:1983if,Andersson:1983ia,Collins:2023cuo}. Thus the total charge of the sector remains exactly $Q_f$, while the bulk detector measures only the complementary charge $Q_f-\langle Q_D\rangle_f$. In this sense, $\langle Q_D\rangle_f$ characterizes a non-perturbative boundary charge assigned to the celestial puncture generated by the defect sector of the future-pointing collinear Wilson line. Its existence represents a possible non-perturbative redistribution of the conserved charge between the observable flow through $S^2\setminus D_\epsilon(\bar n)$ and the puncture sector $D_\epsilon(\bar n)$. If the QCD non-perturbative dynamics respects light-flavor symmetry and loses correlation with the initiating flavor, see Fig.~\ref{fig:hadronization}, one would expect that 
\bea  \label{eq:Qu-qd}
\langle Q_{\rm flow} \rangle_u - \langle Q_{\rm flow}\rangle_d = \langle Q_{\rm flow} \rangle_u + \langle Q_{\rm flow}\rangle_{\bar d} = Q_u - Q_d \,. 
\eea  
More precisely, the relation states that in a Monte Carlo simulation, one should expect to see that the charge difference measured over a $u$-jet and $d$-jet should be identical before or after performing hadronization.  

We note that the renormalization-group evolution of $\langle Q_{\rm flow}\rangle_f$ and $\langle Q_D \rangle_f$ is determined by perturbative ultraviolet dynamics. In this domain, Eq.~\eqref{eq:ward-pert} gives $\langle Q_{\rm flow}\rangle_f^{\rm pert} = Q_f$, so $\langle Q_{\rm flow}\rangle_f$ inherits the vanishing anomalous dimension of the conserved total charge, which then implies that the nonperturbative puncture contribution $\langle Q_D \rangle_f =Q_f- \langle Q_{\rm flow} \rangle_f$ is also renormalization-scale independent.
\bea 
\mu\frac{d}{d\mu} \langle Q_{\rm flow} \rangle_f=0 \,, 
\qquad \mu\frac{d}{d\mu} \langle Q_D \rangle_f =0 \,. 
\eea 
Similar discussion applies to the baryon number $\langle B_{\rm flow} \rangle_f$ and other conserved quantum numbers in QCD.

Independently of the particular hadronization configuration illustrated in Fig.~\ref{fig:hadronization}, confinement imposes a general selection rule on the puncture sector. The anti-fundamental Wilson-line endpoint must be screened by dynamical matter carrying the conjugate color triality. Although the open color index of the Wilson line $W_n$ is contracted with its conjugate in the jet function in Eq.~\eqref{eq:jetQ}, this contraction only makes the squared matrix element color invariant. It does not provide a physical screening excitation in an individual hadronization amplitude. After hadronization, the finite-angle bulk consists of color-singlet hadrons and therefore cannot terminate the nonzero-triality color flux carried by the anti-fundamental Wilson-line endpoint. The endpoint must instead be screened by dynamical quarks and antiquarks bound to it. We call the correlated set of dynamical quarks and antiquarks remaining attached to the Wilson-line endpoint at $\bar n\cdot q=0$ its \emph{screening cloud}. If this cloud contains $k$ endpoint quarks and $\ell$ endpoint antiquarks, color triality~\cite{Greensite:2003bk} requires
\bea 
-1+k-\ell=0\pmod 3, \qquad\Longrightarrow\qquad B_D=\frac{k-\ell}{3}=\frac13+m, \quad m\in \mathbb Z.
\eea 
Here the $-1$ is the color triality of the Wilson line $W_n$, while each quark (antiquark) contributes $+1$ ($-1$). Writing \(N_a=k_a-\ell_a\) for the net number of quarks of flavor \(a\) in the screening cloud, the leading screening possibilities and their associated puncture and  flow quantum numbers are
\bea
\begin{array}{c|c|c|c|c|c}
    \text{screening cloud} & (k,\ell) & B_D & Q_D & B_{\rm flow} & Q_{\rm flow}\\ \hline
    q_a & (1,0) & \dfrac13 & Q_a & 0 & Q_f-Q_a \\[2mm]
    (\bar q_a\bar q_b)_{\mathbf3} & (0,2) & -\dfrac23 & -(Q_a+Q_b) & 1 &Q_f+Q_a+Q_b \\[2mm]
    \text{higher clouds} &\displaystyle \sum_a N_a=1+3m  & \dfrac13+m & \displaystyle\sum_a Q_aN_a & -m & \displaystyle Q_f-\sum_a Q_aN_a
\end{array} \nonumber 
\eea 
Here $Q_a$ is the electric charge of flavor $a$. 

Since additional quark--antiquark excitations from the vacuum incur an energetic cost and are therefore disfavored, we conjecture that the puncture is dominated by the minimal one-quark screening cloud, with the anti-diquark cloud providing the leading correction. Define the conditional endpoint ratio
\bea 
r_{\rm end}\equiv\frac{P[(\bar q\bar q)_{\mathbf 3}\ {\rm screens}\ W_n]}  {P[q\ {\rm screens}\ W_n]}\,, 
\qquad  p_B\equiv \frac{r_{\rm end}}{1+r_{\rm end}} \simeq \frac{r_{qq}}{1+r_{qq}} \,.
\eea 
Here $p_B$ denotes the probability of producing a \((\bar q\bar q)_{\mathbf 3}\) screening cloud. Assuming that vacuum screening probabilities are homogeneous over the celestial sphere, we identify the conditional endpoint ratio $r_{\rm end}$ with the final vacuum diquark-to-quark screening ratio $r_{qq}$.

Neglecting higher screening configurations that are generically suppressed, we then obtain
\bea \label{eq:B-rqq}
\langle B_{\rm flow}\rangle_f \simeq p_B \simeq   \frac{r_{qq}}{1+r_{qq}} \,,
\qquad \langle B_D\rangle_f \simeq \frac13-p_B \,.  
\eea 
Thus the measured jet baryon number directly probes the probability for anti-diquark screening of the Wilson-line puncture. For small $r_{qq}$, we expect substantially suppressed $\langle B_{\rm flow}\rangle_f \simeq r_{qq}$ measured inside a $f$-quark jet, compared with its naive expectation $B_f = 1/3$. 

For the electric charge, let the single screening quark be distributed as
\begin{align}
 {u\bar u}: {d\bar d}: {s\bar s}=1:1:r_s \,.
\end{align}
Its mean quark charge is $\overline Q_q= \sum_a p_aQ_a  = \frac{1-r_s}{3(2+r_s)}$. If the two antiquark flavors in the anti-diquark cloud are drawn independently from the same distribution, then $\overline Q_{\bar q\bar q}=-2\overline Q_q$, and therefore
\bea  
\langle Q_D\rangle_f  \simeq  \overline Q_q \left(1-p_B - 2 p_B \right) \simeq \frac{1-r_s}{2+r_s} \left(\frac13 - \langle B_{\rm flow}\rangle_f \right) \,,
\eea 
which immediately gives 
\bea \label{eq:flow-Q-B}
\langle Q_{\rm flow}\rangle_f \simeq Q_f - \frac{1-r_s}{2+r_s} \left(\frac{1}{3}- \langle B_{\rm flow} \rangle_f\right)\,. 
\eea 
Note that by a similar mechanism, $r_s$ can be probed by the strange number through 
\begin{equation} \label{eq:flow-strangeness-rs}
\langle S_{\rm flow} \rangle_f \simeq S_f + \frac{3 r_s}{2+r_s}\left(\frac13 -\langle B_{\rm flow}\rangle_f \right)  \,, 
\end{equation}
where the $s$-quark carries $S_s = -1$. Eqs.~\eqref{eq:flow-Q-B} and~\eqref{eq:flow-strangeness-rs} predict linear relations among $\langle Q_{\rm flow}\rangle_f-Q_f$, $\langle S_{\rm flow}\rangle_f-S_f$, and $1/3-\langle B_{\rm flow}\rangle_f$. We denote their slopes as $s_{Q{\text -}B}$, $s_{Q{\text -}S}$ and $s_{S{\text -}B}$.

Eventually, we have 
\bea \label{eq:flow-gmn}
\langle Q_{\rm flow}\rangle_f \simeq I_{3,f} + \frac{1}{2}\big(\langle B_{\rm flow} \rangle_f +  \langle S_{\rm flow} \rangle _f \big) \,,
\eea 
which is the Gell-Mann--Nishijima relation~\cite{Nakano:1953zz,Nishijima:1955gxk,Gell-Mann:1956iqa} for the measured jet. Here, $I_{3,f} = Q_f - (S_f+B_f)/2$ is the third component of isospin for the initiating quark. Eq.~\eqref{eq:flow-gmn} should be distinguished from the exact measured-flow identity $\langle Q_{\rm flow}\rangle_f=\langle I_{3,\rm flow}\rangle_f+\tfrac12(\langle B_{\rm flow}\rangle_f+\langle S_{\rm flow}\rangle_f)$, which follows directly by summing the particle-level Gell-Mann--Nishijima relation over the measured region~\cite{Riembau:2024tom}. The nontrivial dynamical content of Eq.~\eqref{eq:flow-gmn} is the initiating-isospin retention relation $\langle I_{3,\rm flow}\rangle_f\simeq I_{3,f}$. Isospin symmetry of the screening cloud makes the contributions from $u$- and $d$-screening configurations cancel on average, yielding $\langle I_{3,D} \rangle_f  \simeq 0$ and hence the stated retention relation.

These relations provide a way to study non-perturbative strangeness and baryon production in QCD using jet measurements and, within the hadronization framework introduced here, to constrain the parameters $r_s$ and $p_B$ (or, approximately, $r_{qq}$). In the next section, we validate these relations using event generators.

\section{Validation against Monte Carlo Simulations}\label{sec:mcvalidation}
We validate our predictions against Monte Carlo simulations using \textsc{Pythia}~8.312~\cite{Bierlich:2022pfr} and \textsc{Herwig}~7.3~\cite{Bellm:2015jjp}, which employ different hadronization models. \textsc{Pythia} uses the Lund string model~\cite{Andersson:1983ia}, whereas \textsc{Herwig} implements the cluster hadronization model~\cite{Webber:1983if}. We consider $e^+e^-$ annihilation and impose a thrust cut, $T > T_{\rm cut} = 0.95$, to select the dijet limit in which the factorization theorem in Eq.~\eqref{eq:fact} applies. Within each selected event, we then measure the conserved quantum numbers flowing into a given hemisphere. We emphasize, however, that our result is not tied to the thrust observable. It applies more generally to any process and measurement, e.g., jets, jet shapes and correlators, for which the factorization theorem in Eq.~\eqref{eq:fact} is valid.
	\begin{figure}[!htb]
		\centering
		\includegraphics[width=0.75\linewidth]{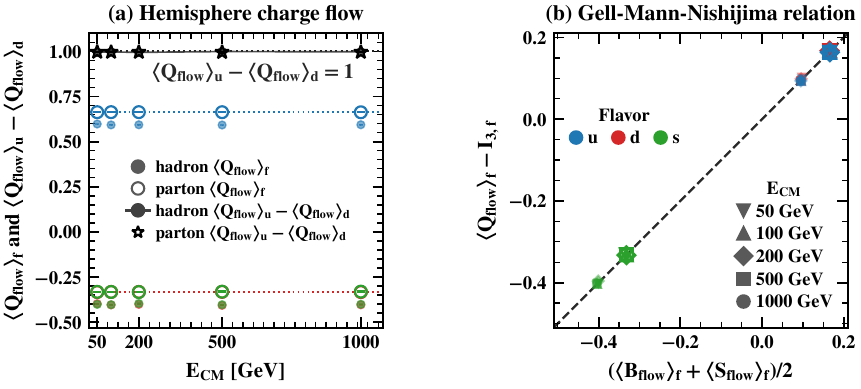}
		\caption{Energy dependence of the measured jet quantum numbers and the jet Gell-Mann--Nishijima relation in $e^+e^- \to q\, \bar q$ events generated with \textsc{PYTHIA}~8.312. Panel (a) shows the flavor-resolved jet charges as functions of $E_{\rm CM}$. The relevant scales are determined by $E_{\rm CM} (1-T_{\rm cut})$. Panel (b) tests the jet Gell-Mann--Nishijima relation for the $u$-, $d$-, and $s$-jets at different center-of-mass energies. In both panels, colors distinguish the initiating flavors, while filled and open markers denote hadron- and parton-level results, respectively.}
		\label{fig:pythia-energy-scan}
	\end{figure}
		
	Fig.~\ref{fig:pythia-energy-scan} shows the quantum-number flow at $E_{\rm CM}=50$, $100$, $200$, $500$, and $1000~\mathrm{GeV}$ measured in \textsc{PYTHIA}~8.312 simulations, restricting the hard process to $u\bar u$, $d\bar d$, and $s\bar s$ production. The requirement $T>0.95$ selects dijet kinematics. The parton-level references are measured at the end of QCD final-state shower (with QED and weak shower off), while the hadron-level results further include hadronization but with decays disabled.
	
	As shown in the left panel, hadronization induces a clear shift in the measured charge for $u$-, $d$-, and $s$-initiated jets.  At the parton level, $\langle Q_{\rm flow}\rangle_f$ reproduces the charge $Q_f$ of the initiating quark, whereas at the hadron level one finds $\langle Q_{\rm flow}\rangle_f\neq Q_f$.  Despite this hadronization correction, the measured charge is nearly independent of $E_{\rm CM}$ at both levels.	Moreover, $\langle Q_{\rm flow}\rangle_u-\langle Q_{\rm flow}\rangle_d$ remains consistent with $Q_u-Q_d$ over the entire energy range, as predicted by Eq.~\eqref{eq:Qu-qd}.
	
\begin{figure}[!htb]
		\centering
		\includegraphics[width=0.75\linewidth]{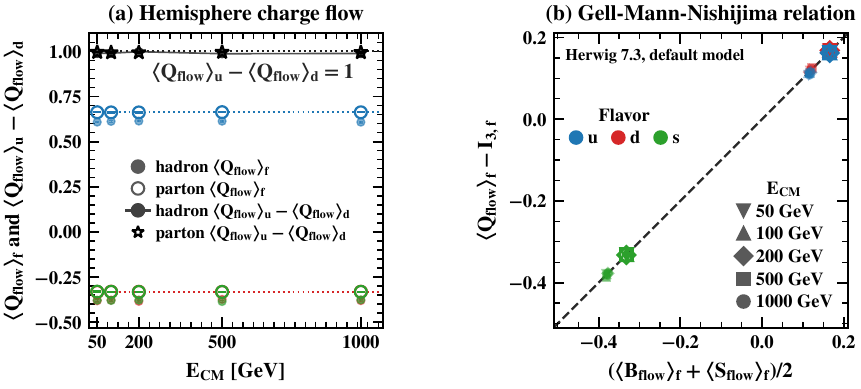}
		\caption{Corresponding energy dependence and jet Gell-Mann--Nishijima tests in $e^+e^- \to q\,\bar q$ events generated with \textsc{Herwig}~7.3 using its default hadronization model.  The panels and marker conventions are defined as in Fig.~\ref{fig:pythia-energy-scan}.}
		\label{fig:herwig-energy-scan}
	\end{figure}

	The right panel of Fig.~\ref{fig:pythia-energy-scan} tests the jet Gell-Mann--Nishijima relation in	Eq.~\eqref{eq:flow-gmn}.  The parton- and hadron-level results both follow the predicted diagonal relation over the full energy scan. Fig.~\ref{fig:herwig-energy-scan} shows that the same behavior is obtained with the default cluster-hadronization model in \textsc{Herwig}.

\begin{figure}[!htb]
		\centering
		\includegraphics[width=\linewidth]{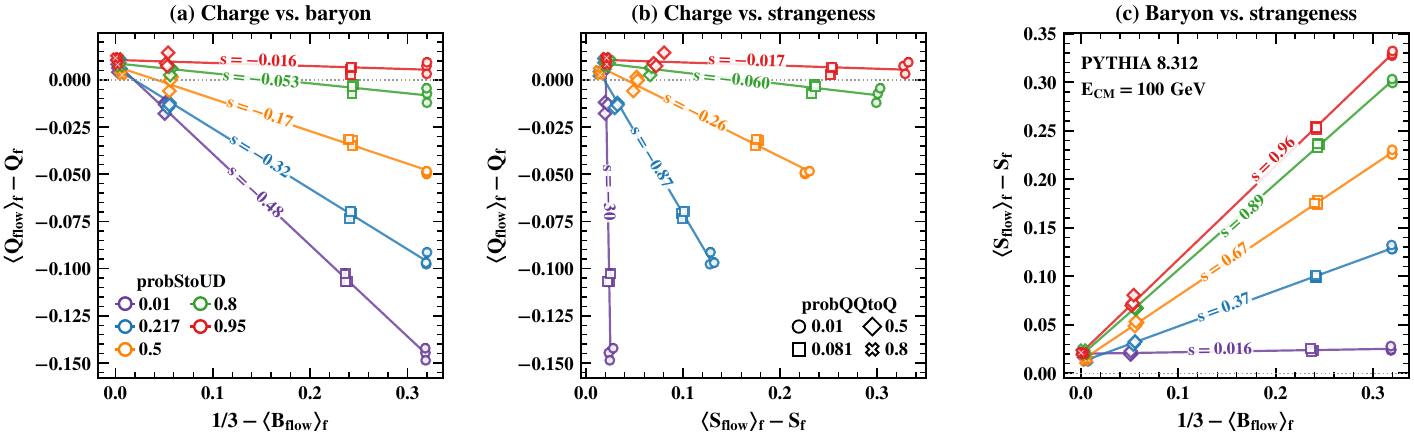}
		\caption{Joint scan of $s$- and $qq$-production parameters in \textsc{PYTHIA}~8.312 at 	$E_{\rm CM}=100~\mathrm{GeV}$ with $T>0.95$. Colors denote 	\texttt{StringFlav:probStoUD}$=0.010,0.217,0.500,0.800,0.950$ (default $0.217$), while marker shapes denote \texttt{StringFlav:probQQtoQ}$=0.010,0.081,0.500,0.800$ (default $0.081$). At fixed \texttt{probStoUD}, each line is a linear fit with a free intercept to the four three-flavor cluster centers; its label gives the fitted slope $s$.}
		\label{fig:pythia-qsb-scan}
	\end{figure}

	Fig.~\ref{fig:pythia-qsb-scan} tests the $Q{\text -}S{\text -}B$ relations by varying two \textsc{PYTHIA} hadronization parameters. The parameter \texttt{StringFlav:probStoUD} controls the relative probability for producing an $s\bar s$ pair at a string breaking, while \texttt{StringFlav:probQQtoQ} controls the relative rate of	diquark--antidiquark production. 

    We take $E_{\rm CM}=100~\mathrm{GeV}$ as our default scattering energy in Fig.~\ref{fig:pythia-qsb-scan}. Here we do not distinguish the $u$-, $d$-, and $s$-initiated points because, for fixed \textsc{Pythia} parameters, they are nearly indistinguishable and form a compact cluster in all panels.
	
	For each fixed value of \texttt{probStoUD}, the clusters organize into $5$ nearly linear families. We perform fits with free intercepts to them and compare the values of $r_s$ extracted from the fitted slopes. \Cref{tab:pythia-slope-rs} shows the numerical values with  uncertainties propagated from the slope fits. We note that the \textsc{PYTHIA} string model contains additional non-perturbative hadronization settings, nevertheless, we get good agreement $r_s \approx \texttt{probStoUD}$. Furthermore, the three projections yield mutually consistent values of $r_s$ within the precision of the scan, supporting the predictions in Eqs.~\eqref{eq:flow-Q-B} and~\eqref{eq:flow-strangeness-rs}.

	\begin{table}[!htb]
		\centering
		\caption{Fitted slopes and the corresponding effective strange-pair ratios for the \textsc{PYTHIA} scan in Fig.~\ref{fig:pythia-qsb-scan}. Each $r_s$ value is obtained from the corresponding fitted slope.  Parentheses give the one-standard-error	uncertainty in the final quoted digits.}
		\label{tab:pythia-slope-rs}
		\begingroup	\small \setlength{\tabcolsep}{4pt} \renewcommand{\arraystretch}{1.25}
		\begin{tabular}{lcccccc}
			\toprule
			\texttt{probStoUD} & $s_{Q{\text -}B}$ & $s_{Q{\text -}S}$ & $s_{S{\text -}B}$ & $r_s^{Q{\text -}B}$ & $r_s^{Q{\text -}S}$ & $r_s^{S{\text -}B}$ \\
			\midrule
			$0.010$ & $-0.4839(44)$ & $-29.8(17)$ & $0.0161(10)$ & $0.0217(59)$ & $0.01105(61)$ & $0.01080(69)$ \\
			$0.217$ & $-0.3197(77)$ & $-0.872(27)$ & $0.3666(29)$ & $0.273(13)$ & $0.2766(61)$ & $0.2785(25)$ \\
			$0.500$ & $-0.1721(70)$ & $-0.256(11)$ & $0.6718(28)$ & $0.560(15)$ & $0.566(11)$ & $0.5771(31)$ \\
			$0.800$ & $-0.0531(24)$ & $-0.0599(28)$ & $0.8859(42)$ & $0.8489(64)$ & $0.8477(61)$ & $0.8381(57)$ \\
			$0.950$ & $-0.0161(58)$ & $-0.0167(61)$ & $0.9644(74)$ & $0.952(17)$ & $0.952(16)$ & $0.948(11)$ \\
			\bottomrule
		\end{tabular}
		\endgroup
	\end{table}
	
Panels (a) and (c) of Fig.~\ref{fig:pythia-qsb-scan} provide a complementary test of the baryon-number dependence.  At fixed \texttt{probQQtoQ}, the points corresponding to different \texttt{probStoUD} values are nearly vertically aligned.  Consequently, $\langle B_{\rm flow}\rangle_f$ is largely independent of the strange-pair parameter, consistent with our prediction.   Averaging over the $5$ \texttt{probStoUD} settings and $3$ initiating flavors yields $\langle B_{\rm flow}\rangle_f=0.0139$, $0.0922$, $0.2790$, and $0.3298$ for $\texttt{probQQtoQ}=0.010$, $0.081$, $0.500$, and $0.800$, respectively. Eq.~\eqref{eq:B-rqq} then gives $r_{qq}\simeq 0.0141$, $0.1016$, $0.3870$, and $0.4921$. These ratios should not be identified with \texttt{probQQtoQ}, which weights diquark production only at eligible steps of the state-dependent fragmentation history. Once a diquark is selected, subsequent steps are constrained, so this local input weight need not equal the final ratio. As a complementary event-level check, we find that $P(B_{\rm flow}=1)/P(B_{\rm flow}=0)$ agrees with   the inferred $r_{qq}$ values within $10\%$, see Table~\ref{tab:topology-check}, supporting the two-channel picture and the extraction in Eq.~\eqref{eq:B-rqq}.
	
As a generator-level cross-check, we repeat the scan with the cluster hadronization model in \textsc{Herwig}.  The default  \texttt{Baryonic} color reconnection used in Fig.~\ref{fig:herwig-energy-scan} does not provide an adjustable baryon-number direction through \texttt{MesonToBaryonFactor}. For the joint scan we therefore use the \texttt{BaryonicMesonic} model, in which \texttt{MesonToBaryonFactor} controls the relative competition between mesonic and baryonic reconnection topologies.  The parameters \texttt{SplitPwtSquark} and \texttt{PwtSquark} control strange-quark production during nonperturbative gluon splitting and during cluster fission and decay, respectively.
	
	\begin{figure}[!htb]
		\centering
		\includegraphics[width=\linewidth]{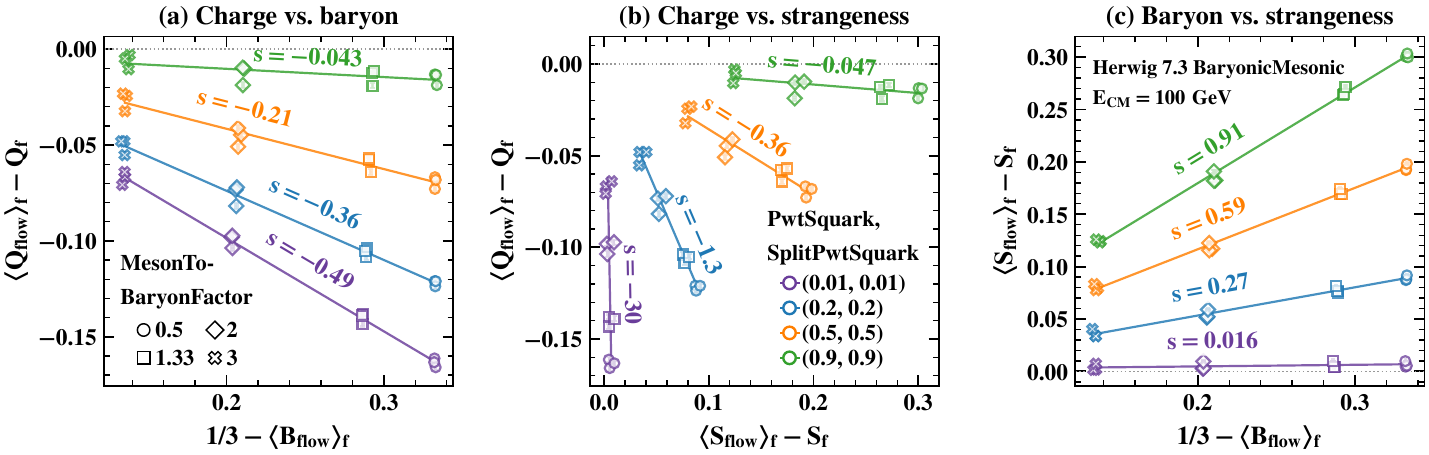}
		\caption{Corresponding joint scan of strange- and baryon-production parameters using the \textsc{Herwig}~7.3	\texttt{BaryonicMesonic} cluster-hadronization model at $E_{\rm CM}=100~\mathrm{GeV}$ with $T>0.95$. Colors denote the matched $(\texttt{PwtSquark},\texttt{SplitPwtSquark})$ settings $(0.01,0.01)$, $(0.2,0.2)$, $(0.5,0.5)$, and $(0.9,0.9)$, while marker shapes denote \texttt{MesonToBaryonFactor}$=0.5,1.333,2,3$. Each parameter point contains $1.5\times10^6$ generated events. The panels and fits are defined as in Fig.~\ref{fig:pythia-qsb-scan}, and each line label gives its directly fitted slope $s$.}
		\label{fig:herwig-qsb-scan}
	\end{figure}
	
	The corresponding slopes and effective values of $r_s$ are summarized in Table~\ref{tab:herwig-slope-rs}.  Under these settings, \textsc{Herwig} exhibits the same linear relations, the same weak dependence of $\langle B_{\rm flow}\rangle_f$ on the strange-pair parameters, and the same consistent determination of $r_s$ from the three projections as \textsc{PYTHIA}, despite the different microscopic hadronization model.

	\begin{table}[!htb]
		\centering
		\caption{Fitted slopes and the corresponding effective strange-pair ratios for the \textsc{Herwig} scan in Fig.~\ref{fig:herwig-qsb-scan}. Each $r_s$ value is obtained from the corresponding fitted slope. Parentheses give the one-standard-error uncertainty in the final quoted digits.}
		\label{tab:herwig-slope-rs}
		\begingroup \small \setlength{\tabcolsep}{4pt} \renewcommand{\arraystretch}{1.25}
		\begin{tabular}{lcccccc}
			\toprule
			$\substack{\texttt{PwtSquark}\\=\texttt{SplitPwtSquark}}$ & $s_{Q{\text -}B}$ & $s_{Q{\text -}S}$ & $s_{S{\text -}B}$ & $r_s^{Q{\text -}B}$ & $r_s^{Q{\text -}S}$ & $r_s^{S{\text -}B}$ \\
			\midrule
			$0.01$ & $-0.4868(48)$ & $-30.1(45)$ & $0.0155(21)$ & $0.0177(65)$ & $0.0110(16)$ & $0.0104(14)$ \\
			$0.2$  & $-0.3607(20)$ & $-1.337(20)$ & $0.2697(47)$ & $0.2047(33)$ & $0.1996(23)$ & $0.1975(38)$ \\
			$0.5$  & $-0.210(13)$ & $-0.356(32)$ & $0.587(16)$ & $0.479(27)$ & $0.483(23)$ & $0.486(16)$ \\
			$0.9$  & $-0.043(13)$ & $-0.047(15)$ & $0.914(25)$ & $0.876(35)$ & $0.878(34)$ & $0.876(34)$ \\
			\bottomrule
		\end{tabular}
		\endgroup
	\end{table}

\section{Conclusion}\label{sec:conclusion}
In this note, we have established that the shift in the joint jets/flows of electric charge, baryon number, and strangeness relative to the corresponding quantum number of the initiating parton directly encodes the information on QCD vacuum hadronization. At the perturbative level, each flow moment equals the charge of the initiating parton; any shift observed after hadronization therefore measures the nonperturbative defect charge at the Wilson-line endpoint. Because the underlying currents are conserved, both the flow and defect first moments are renormalization-scale independent at leading power. Gauge invariance fixes the total triality of the Wilson line and its screening cloud to zero, sharply restricting the quark content and the quantum numbers of the screening cloud.  Interpreting these shifts in terms of vacuum hadronization probabilities then requires only generic dynamical assumptions: the minimal one-quark and anti-diquark screening clouds dominate and the screening flavors lose memory of the initiating flavor and are sampled approximately independently from a common light-flavor distribution. None of these assumptions prescribes a specific string or cluster algorithm.

Within this framework, the measured baryon number $\langle B_{\rm flow} \rangle_f$ determines the final anti-diquark-to-quark endpoint ratio $r_{qq}$, predicting a near-vanishing net jet baryon number despite the initiating-quark value $B_f=1/3$, if the vacuum excitation to anti-diquark-to-quark is suppressed and $r_{qq} \ll 1$. The same mechanism predicts an $r_s$-dependent strangeness shift correlated with the baryon deficit  $1/3-\langle B_{\rm flow}\rangle_f$. Consequently, $u$- and $d$-initiated jets, although carrying $S_f = 0$, acquire a nonzero mean net strangeness. The three pairwise relations among the average charge, baryon-number, and strangeness flows then provide mutually consistent determinations of the strange-to-light ratio $r_s$.
      
A more general consequence is the jet Gell-Mann--Nishijima relation, $\langle Q_{\rm flow}\rangle_f\simeq I_{3,f}+[\langle B_{\rm flow}\rangle_f+ \langle S_{\rm flow}\rangle_f]/2$. This closure relation only follows from the QCD flavor symmetry in the screening cloud. The joint flows therefore provide both a quantitative probe of vacuum screening and a broadly model-insensitive consistency test of flavor and charge transport during hadronization.

The framework extends naturally to heavy-flavor flows. Since nonperturbative vacuum excitation of heavy flavor pairs is strongly suppressed, the net charm and bottomness of a heavy-flavor jet should retain those of the initiating quark. The remaining charge flows follow the same light-flavor screening as in light jets, implying $\langle {\cal Q}_{\rm flow}\rangle_c\simeq\langle {\cal Q}_{\rm flow }\rangle_u$ and $\langle {\cal Q}_{\rm flow}\rangle_b\simeq\langle {\cal Q}_{\rm flow }\rangle_d$ for \({\cal Q}= Q,B,S\). They must also satisfy the generalized jet Gell-Mann--Nishijima relation including charm and bottomness.

Initial studies with Monte Carlo event generators reproduce the predicted charge-flow relations and recover the corresponding effective screening probabilities, demonstrating that the defect charge encodes fragmentation dynamics beyond those accessible through inclusive hadron yields. The theoretical structure is also insensitive to infrared cutoffs. We find that imposing a fiducial momentum cut of order $\Lambda_{\mathrm{QCD}}$ does not modify the underlying scale invariance. We expect that future experimental measurements will provide stringent tests of these predictions and assess the potential of probing vacuum hadronization. Relevant experimental capabilities have been demonstrated in $e^+e^-$~\cite{DELPHI:1995dso,DELPHI:1996nbj,DELPHI:2000kzb,SLD:1998coh,OPAL:1998izp}, and the measurements at RHIC and the LHC~\cite{ALICE:2008ngc,ALICE:2016zzl,STAR:2008med,ALICE:2021vxl,ALICE:2024iqc,ALICE:2021hjb,ALICE:2023asw,ALICE:2019nbs}. 

Although our discussion primarily concerns hadronization in the vacuum, extending these measurements across collision systems, from high-multiplicity $pp$ and $pA$ collisions to heavy-ion collisions ($AA$), would be particularly interesting. In hadronic and nuclear collisions, the factorization boundary may be screened by additional mechanisms through color reconnection, interactions with the surrounding matter, or recombination with medium partons~\cite{Christiansen:2015yqa,Fries:2003vb,Greco:2003xt,Fries:2008hs,Han:2016uhh}. The beam-remnant effects may also induce long-range quantum-number transport through baryon-junction mechanisms, which lead to non-trivial correlation between $Q$ and $B$~\cite{KHARZEEV1996238,Christiansen:2015yqa,STAR:2024lvy}. These mechanisms may modify the effective relative weights of light-quark, strange-quark, and baryonic screening channels. In nuclear collisions, the formation of a thermal partonic environment may suggest that, with increasing system size or multiplicity, the redistribution of quantum numbers between the measured flow and the defect sector can gradually approach a thermodynamic limit. This expectation is qualitatively motivated by the observed strangeness enhancement and baryon-to-meson enhancement at intermediate transverse momentum in high-multiplicity $pp$ and heavy-ion collisions~\cite{Rafelski:1982pu,Andronic:2017pug,STAR:2006uve,STAR:2007cqw,ALICE:2013mez,ALICE:2016fzo,ALICE:2018pal}. Therefore, studying these jet quantum-number flows for jets propagating through different collision systems would provide the opportunity to directly probe how vacuum fragmentation is modified by the broader range of hadronization mechanisms that exist in the medium.

\paragraph*{Acknowledgments.---}
We thank Hao Chen, Zhongbo Kang, Kyle Lee, Ian Moult, Manqi Ruan, Zhen Xu and Feng Yuan for discussions. 
W.~K. is supported by National Natural Science Foundation of China under Grant No. 12575140.
H.~T.~L. is supported by the National Natural Science Foundation of China under Grant No. 12275156, No. 12321005 and the Shandong Provincial Department of Science and Technology under Project No. TQ012025001. 
W.~L. and D.~Y.~S. are supported by the National Natural Science Foundation of China under Grant No.~12275052, No.~12147101, No.~12547102, and the Innovation Program for Quantum Science and Technology under grant No. 2024ZD0300101.
W.~L. is also supported by the China Postdoctoral Science Foundation under grant No. 2025M783369.
X.~L. is supported by the National Natural Science Foundation of China under Grant No.~12547109 and Fundamental Research Funds for the Central Universities, Beijing Normal University.
X.~L. and D.~Y.~S. would like to thank the Erwin-Schr\"odinger International Institute for Mathematics and Physics at the University of Vienna for partial support during the Programme `New Paradigms for Harnessing Quantum Field Theory at Colliders', July 27 - August 28, 2026. 

\bibliographystyle{apsrev4-1}
\bibliography{refs.bib}

\clearpage
\appendix
\onecolumngrid

\makeatletter
\@removefromreset{equation}{section}
\makeatother

\setcounter{equation}{0}
\setcounter{figure}{0}
\setcounter{table}{0}
\renewcommand{\theequation}{S-\arabic{equation}}
\renewcommand{\thefigure}{S-\arabic{figure}}
\renewcommand{\thetable}{S-\arabic{table}}

\renewcommand{\theHequation}{supp.equation.\arabic{equation}}
\renewcommand{\theHfigure}{supp.figure.\arabic{figure}}
\renewcommand{\theHtable}{supp.table.\arabic{table}}

\allowdisplaybreaks

\section*{Appendices} 

\section{Flavor ratios of the one-point correlator} \label{app:opcc}
\begin{figure}[!htb]
		\centering
		\includegraphics[width=0.98\linewidth]{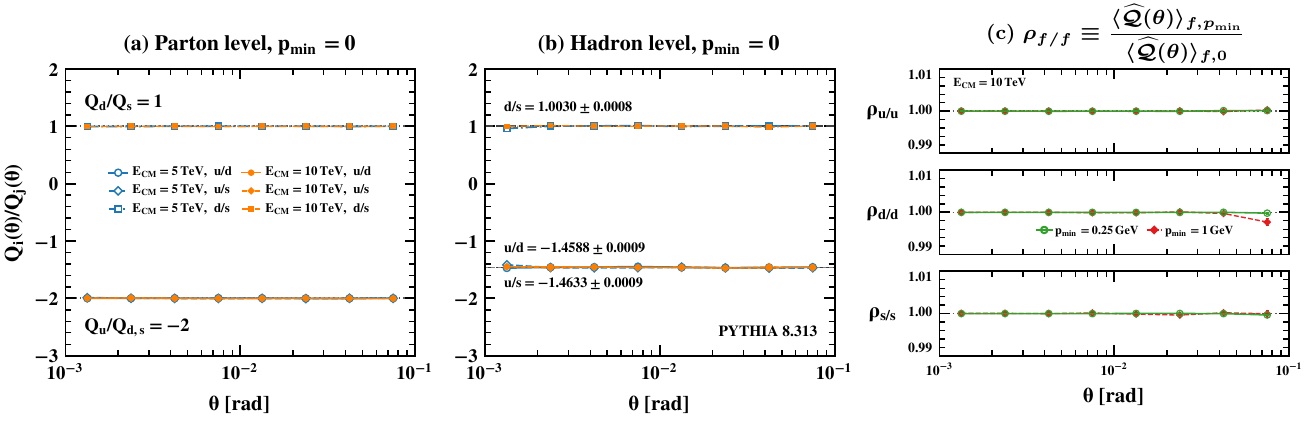}
		\caption{Flavor ratios of the OPCC in $e^+e^-\to\gamma^*\to f\bar f$ events generated with	\textsc{PYTHIA}~8.312 at $E_{\rm CM}=5$ and $10~\mathrm{TeV}$. Panels (a) and (b) show the parton- and hadron-level results, respectively. Panel (c) shows the hadron-level same-flavor response at $E_{\rm CM}=10~\mathrm{TeV}$ for minimum hadron momentum acceptance $p_{\min}=0.25$ and $1~\mathrm{GeV}$ relative to $p_{\min}=0$. The $5~\mathrm{TeV}$ samples contain $3\times10^7$ events per initiating flavor, and the $10~\mathrm{TeV}$ samples contain $2\times10^7$ events per flavor. }
		\label{fig:opcc-flavor-ratios}
	\end{figure}

We verify that our results extend to correlators defined in terms of conserved quantum numbers. As a concrete example, in Fig.~\ref{fig:opcc-flavor-ratios}, we consider the one-point charge correlator (OPCC)~\cite{Riembau:2024tom,Cao:2026fzq,Li:2026fhd, Barata:2026eth, Cao:2026kcg, Zhang:2026vdo,Zhang:2026esd,Zhang:2026emt}, which probes the net charge flow along a given direction $\langle \widehat{Q}(\Omega) \rangle $. We follow~\cite{Li:2026fhd} to measure the OPCC within a parton-$f$ initiated jet, in which Ref.~\cite{Li:2026fhd} shows that at the leading power for $\theta \ll 1$, 
\bea 
\frac{ \langle \widehat{Q}(\Omega) \rangle_{f_1} }{\langle \widehat{Q}(\Omega) \rangle_{f_2}} \simeq 
\frac{\langle Q_{\rm flow} \rangle_{f_1} }{\langle Q_{\rm flow} \rangle_{f_2} }\,,
\eea 
independent of the direction. Following Eq.~\eqref{eq:ward-pert}, at the parton level, the ratio reduces to 
\bea 
\frac{\langle Q_{\rm flow} \rangle_{f_1}^{\rm pert} }{\langle Q_{\rm flow} \rangle_{f_2}^{\rm pert} } = \frac{Q_{f_1}}{Q_{f_2}} \,. 
\eea 

	In \cref{fig:opcc-flavor-ratios}, we verify this prediction at two center-of-mass energies.  At the parton level, the ratios do follow the initiating-quark charges, $\langle \widehat {Q}(\theta) \rangle_u/\langle \widehat {Q}(\theta) \rangle_d= \langle \widehat {Q}(\theta) \rangle_u/\langle \widehat {Q}(\theta) \rangle_s=-2$ and $\langle \widehat {Q}(\theta) \rangle_d/\langle \widehat {Q}(\theta) \rangle_s=1$, throughout the plotted range.
	
	As discussed in the main text, hadronization shifts the charge away from the corresponding initiating-quark charge, which also modifies the ratio. The corresponding OPCC ratios, obtained by combining the $5$ and $10~\mathrm{TeV}$ samples, are
	\begin{equation}
		\frac{\langle Q_{\rm flow}\rangle_u }{\langle Q_{\rm flow}\rangle_d}\simeq -1.4635(10),
		\qquad \frac{\langle Q_{\rm flow}\rangle_u}{\langle Q_{\rm flow}\rangle_s}\simeq -1.4680(10),
		\qquad \frac{\langle Q_{\rm flow}\rangle_d}{\langle Q_{\rm flow}\rangle_s}\simeq 1.0031(8).
		\label{eq:opcc-pythia-ratios}
	\end{equation}

As a comparison, we quote the $E_{\rm CM}=1~\mathrm{TeV}$ hemisphere charge result from Fig.~\ref{fig:pythia-energy-scan}, which gives
	\begin{align}
		\langle Q_{\rm flow}\rangle_u &= 0.5940(12), 
        &\langle Q_{\rm flow}\rangle_d &= -0.4039(25), 
        &\langle Q_{\rm flow}\rangle_s &= -0.4009(25),	\nonumber\\
		\frac{\langle Q_{\rm flow}\rangle_u}{\langle Q_{\rm flow}\rangle_d} & = -1.471(10),
        & \frac{\langle Q_{\rm flow}\rangle_u}{\langle Q_{\rm flow}\rangle_s} & = -1.482(10), 
        & \frac{\langle Q_{\rm flow}\rangle_d}{\langle Q_{\rm flow}\rangle_s} & = 1.007(9)\,, 
		\label{eq:inclusive-flow-ratios}
	\end{align}
	in agreement with the OPCC measurement. This agreement shows that the result is not tied to a particular observable definition, but applies more generally to correlators of conserved quantum-number flows. Panel~(c) of \mbox{\cref{fig:opcc-flavor-ratios}} compares the hadron-level OPCC for $p_{\min}=0.25$ and $1~\mathrm{GeV}$ with the uncut result at $E_{\rm CM}=10~\mathrm{TeV}$ through $\rho_{f/f}(\theta;p_{\min})=\langle \widehat {Q}_f(\theta)\rangle_{p_{\min}}/\langle \widehat{Q}_f(\theta)\rangle_0$. It is consistently unity and shows that fiducial momentum selection leaves the OPCC angular profile essentially unchanged.

\begin{figure}[!htb]
		\centering
		\includegraphics[width=0.98\linewidth]{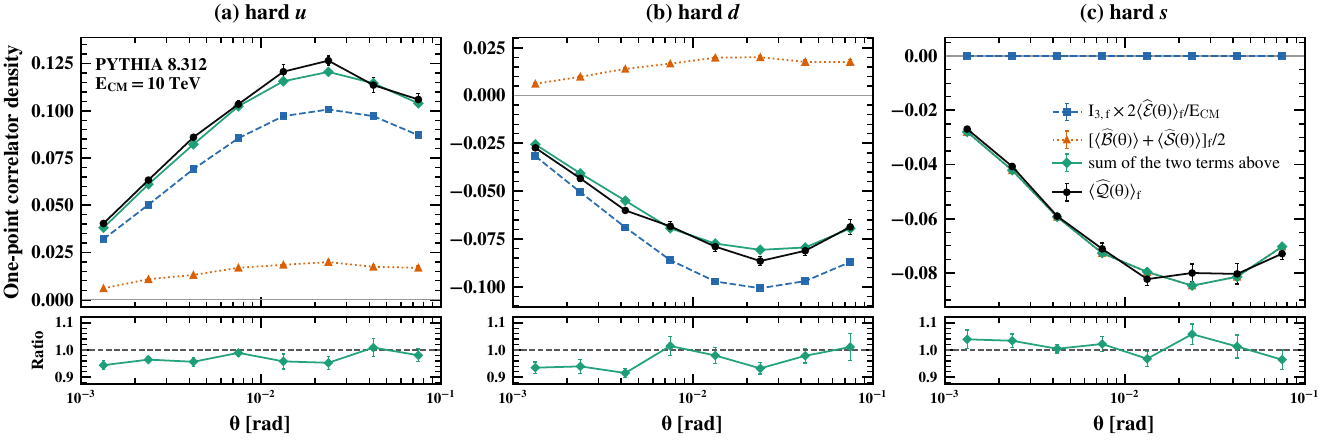}
		\caption{Comparison of the one-point correlators that contribute to \cref{eq:Correlator_GMN} for $u$, $d$, $s$ quark initiated jets at hadron level respectively. No thrust cut was imposed. The lower panels show the ratio $\Big[ I_{3,f}\,\frac{\langle \widehat{\cal E}(\theta)\rangle_f}{E_{\rm CM}/2} + \frac{\langle \widehat{\cal B}(\theta)\rangle_f +\langle \widehat{\cal S}(\theta)\rangle_f}{2}\Big]\Big/ \langle \widehat{\cal Q}(\theta)\rangle_f$, i.e., the two sides of the \cref{eq:Correlator_GMN}. The events are generated with \textsc{PYTHIA}~8.312 for $e^+e^-\to q\bar q$ at $E_{\rm CM}=10~\mathrm{TeV}$. }
		\label{fig:opcc-gmn-decomposition}
	\end{figure}

We further test the Gell-Mann--Nishijima relation bin by bin in $\theta$ in Fig.~\ref{fig:opcc-gmn-decomposition}. To properly normalize the correlators, we approximate the normalization factor by the one-point energy correlator, which agrees with the normalization of the one-point correlators up to LL accuracy 
\bea \label{eq:opcc-gmn-decomposition} 
    &   \quad  \frac{\langle \widehat{Q}(\theta)\rangle_f- \frac{1}{2}[\langle \widehat{B}(\theta)\rangle_f+\langle \widehat{S}(\theta)\rangle_f]} { 2\langle \widehat{\cal E}(\theta)\rangle_f/E_{\rm CM}}
    \simeq \langle Q_{\rm flow} \rangle_f - \frac{1}{2} [\langle B_{\rm flow} \rangle_f+ \langle S_{\rm flow} \rangle_f] \simeq I_{3,f}\, \\ 
     \Longrightarrow    & \quad \langle \widehat{Q}(\theta)\rangle_f \simeq I_{3,f}\,\frac{\langle \widehat{\cal E}(\theta)\rangle_f}{E_{\rm CM}/2}	+		\frac{\langle \widehat{B}(\theta)\rangle_f	+\langle \widehat{S}(\theta)\rangle_f}{2}\,,\label{eq:Correlator_GMN}
\eea 
and is therefore sufficient for the present analysis. We leave a precise calculation of the normalization factor to future work. 

Fig.~\ref{fig:opcc-gmn-decomposition} shows that the correlators closely follow the relation in Eq.~\eqref{eq:opcc-gmn-decomposition}. This provides a differential test of the jet Gell-Mann--Nishijima relation and confirms that our formalism extends to correlator observables.

   \section{Event-level test of the minimal screening approximation} \label{app:topology-check}
   The extraction of \(r_{qq}\) from the first moment \(\langle B_{\rm flow}\rangle_f\) assumes that the Wilson-line endpoint is   predominantly screened either by a single quark or by an anti-diquark. These two minimal screening clouds carry defect baryon numbers \(B_D=1/3\) and \(-2/3\), respectively, and therefore correspond to \(B_{\rm flow}=0\) and \(1\) in a quark-initiated hemisphere. To test this two-channel truncation, we retain the event-by-event distribution
   \begin{equation}
     P_m \equiv P(B_{\rm flow}=m), \qquad \sum_m P_m=1.
     \label{eq:Bm-distribution}
   \end{equation}
   The associated topology estimator is defined as
   \begin{equation} \label{eq:r-topology}
     r_{\rm topology}\equiv\frac{P_1}{P_0}. 
   \end{equation}
Here we have assumed the homogeneous-vacuum identification adopted in the main text, to identify \(r_{qq}\) with \(r_{\rm end}\).
 
Table~\ref{tab:topology-check} compares \(r_B=\langle B_{\rm flow}\rangle/(1-\langle B_{\rm flow}\rangle)\) with   \(r_{\rm topology}\). At each \texttt{probQQtoQ} value, the probabilities are evaluated from the same selected event sample and combined over the $5$ \texttt{probStoUD} settings and three initiating flavors used in the main analysis. The two estimators agree to within  $9\%$ over the full scan. This agreement supports the use of the minimal screening approximation, while the systematic difference between the estimators quantifies its residual correction.
  \begin{table}[!hbt]
     \centering
     \caption{Event-level test of the two-channel screening approximation. Here \(r_B=\langle B_{\rm flow}\rangle/(1-\langle B_{\rm flow}\rangle)\), \(r_{\rm topology}=P(B_{\rm flow}=1)/P(B_{\rm flow}=0)\). Agreement between \(r_B\) and \(r_{\rm topology}\) validates the minimal screening assumption.}
     \label{tab:topology-check}
     \begin{tabular}{c c c }
			\toprule
			\texttt{probQQtoQ} \,\, & \(\quad r_{qq} \simeq r_B \quad \) & \(\quad r_{\rm topology} \quad \)
            \\ \midrule
			0.010 & \(0.01410(9)\) & \(0.01526(9)\) \\
			0.081 & \(0.1015(3)\)  & \(0.1102(3)\)  \\
			0.500 & \(0.3869(7)\)  & \(0.4215(6)\) \\
			0.800 & \(0.4921(8)\)  & \(0.5341(8)\)
            \\ \bottomrule
		\end{tabular}
   \end{table}

\medskip

\end{document}